\documentclass[aps,prd,nofootinbib,showpacs,amssymb,twocolumn]{revtex4}
\usepackage{amsmath,amsfonts}
\usepackage{epsfig,subfigure}
\usepackage{ifpdf}
\usepackage{amsmath}
\usepackage{graphicx}
\usepackage{color}
\usepackage{natbib}
\usepackage{multirow}

\newcommand{\be}{\begin{equation}}
\newcommand{\ee}{\end{equation}}
\newcommand{\bea}{\begin{eqnarray}}
\newcommand{\eea}{\end{eqnarray}}

\newcommand\hidetosubmit[1]{}
\renewcommand\hidetosubmit[1]{#1}
\newcommand\optional[1]{}
\newcommand\ForInternalReference[1]{}

\newcommand\unitVec[1]{\hat{\mathbf{#1}}}
\newcommand\unitVector[1]{\hat{\mathbf{#1}}}
\newcommand\mc{ {{\cal M}_c}}

\newcommand\Y[1]{Y^{(#1)}}

\newcommand\Like{{\cal L}}

\newcommand\qmstateproduct[2]{\left<#1|#2\right>}

\newcommand\secRO[1]{\noindent {\it #1}. -- }

\newcommand\abbrvILE{PBOO}

\newcommand\citeMCMC{\cite{LIGO-CBC-S6-PE,2011PhRvD..83h2002D,2011PhRvD..84f2003C,gr-extensions-tests-Europeans2011,gwastro-mergers-PE-Aylott-LIGOATest,2011ApJ...739...99N,2012PhRvD..85j4045V,gw-astro-PE-Raymond,gw-astro-PE-lalinference-v1,gwastro-em-First2Years-2014,2014PhRvL.112y1101V,gwastro-mergers-PNLock-Distinguish-Daniele2015,2015ApJ...798L..17C,gwastro-PE-Salvo-EvidenceForAlignment-2015,gwastro-pe-Tyson-AstroSample-MassGap2015,2015arXiv150606032M}}
\newcommand\citeTideGWBasics{\cite{2009PhRvD..79l4033R,2009PhRvD..79l4032R,gwastro-ns-TidesDoWork-Hinderer2009,gwastro-ns-EOSViaBHNSMergers-Lackey2011,2012PhRvD..85l3007D,2014PhRvD..89b4031D,2015PhRvD..91d3002L,2013PhRvD..88j4040M,2014PhRvD..89b1303Y,2015PhRvD..92b3012A}}
\newcommand\citeAstroConstraints{\cite{gwastro-Ilya-ConfProc-NRDA-2010,2010CQGra..27k4007M,PSconstraints3-MassDistributionMethods-NearbyUniverse,2013PhRvD..87j4028G,popsyn-LowMetallicityImpact2c-StarTrackRevised-2014,popsyn-InterpretStarTrack-LIGO-FairhurstOhme-2015,2015MNRAS.450L..85M}}

\newcommand\citeComplexOverlap{\cite{gwastro-mergers-nr-Alignment-ROS-Methods,gwastro-mergers-nr-Alignment-ROS-IsJEnough,gwastro-mergers-nr-Alignment-ROS-CorotatingWaveforms,gwastro-mergers-nr-Alignment-ROS-Polarization,gwastro-mergers-HeeSuk-FisherMatrixWithAmplitudeCorrections}}

\begin{document}

\author{Richard O'Shaughnessy}
\email{oshaughn@mail.rit.edu}
\affiliation{Center for Computational Relativity and Gravitation, Rochester Institute of Technology, NY 14623, USA}

\author{Prakash Nepal}
\affiliation{University of Wisconsin-Milwaukee, Milwaukee, WI 53211, USA}

\author{A. Lundgren}
\affiliation{Albert-Einstein-Institut, Callinstr. 38, 30167 Hannover, Germany}

\title{A semianalytic Fisher matrix for precessing  binaries with a single significant spin }
\date{\today}

\begin{abstract}
Gravitational waves from a binary with a single dynamically significant spin, notably including  precessing black
hole-neutron star (BH-NS) binaries,  let us constrain that binary's properties:
the two masses and the dominant black hole spin.    Based on a straightforward  fourier transform of $h(t)$ enabled by the corotating
frame, we show the  Fisher matrix for precessing binaries can be well-approximated by an extremely simple semianalytic
approximation.  This approximation can be easily
understood as a weighted average of independent information channels, each associated with one gravitational wave harmonic.
Generalizing previous studies of  nonprecessing binaries to include critical symmetry-breaking precession effects
required to understand plausible astrophysical sources,  our ansatz can be applied to address how well
gravitational wave measurements can address a wide range of  astrophysical and fundamental questions.  Our
approach provides a simple method to assess  what parameters gravitational wave detectors can measure, how
well, and why.
\end{abstract}

\maketitle
\secRO{Introduction}
Ground-based instruments like LIGO  \cite{2015CQGra..32g4001T} and Virgo
\cite{TheVirgo:2014hva})  will soon identify and measure the properties   \citeMCMC{} of the relatively well-understood
gravitational wave (GW) signal 
from the nearly adiabatic and quasicircular inspiral of the lowest-mass coalescing compact binaries (CBCs)
\cite{2003PhRvD..67j4025B,2004PhRvD..70j4003B,2004PhRvD..70f4028D,BCV:PTF,2005PhRvD..71b4039K,2005PhRvD..72h4027B,2006PhRvD..73l4012K,2007MNRAS.374..721T,2008PhRvD..78j4007H,gr-astro-eccentric-NR-2008,gw-astro-mergers-approximations-SpinningPNHigherHarmonics,gw-astro-PN-Comparison-AlessandraSathya2009}: 
binaries consisting of either black holes or neutron stars with total
masses $M=m_1+m_2\le 16 M_\odot$ and intrinsic spins $\mathbf{S}_1,\mathbf{S}_2$ that satisfy the Kerr limit
$|\mathbf{S}_i|/m_i^2\le 1$.   
These measurements' accuracy  determines the range of astrophysical and fundamental
questions that can be addressed via gravitational waves, including but not limited to identifying
 how coalescing compact binaries form \citeAstroConstraints{};
how the universe expands \cite{2005ApJ...629...15H};
how high-density nuclear matter behaves and responds  \citeTideGWBasics{};
and even how reliably general relativity describes the inspiral, coalescence, and gravitational radiation from each
event \cite{2015arXiv150107274B}.  
In general, astrophysical formation channels \cite{clusters-2005,2004PhRvD..69j2002G,2007PhR...442...75K,2008ApJ...682..474B,2010ApJ...719L..79F,2013PhRvD..87j4028G,2015PhRvL.115e1101R}  will populate generic spin
orientations, not just   high-symmetry, nonprecessing configurations with
$\mathbf{S}_1,\mathbf{S}_2$ parallel to $\mathbf{L}$.  For these most likely sources,   spin-orbit and spin-spin couplings cause the
misaligned angular momenta to precess
\cite{ACST,gwastro-mergers-PNLock-Morphology-Kesden2014,gwastro-mergers-PNLock-PRLFollowup}, breaking degeneracies
present in the high-symmetry case and thus enabling higher-precision measurements
\cite{2004PhRvD..70d2001V,2006PhRvD..74l2001L,2011PhRvD..84b2002L,gw-astro-PE-Raymond,LIGO-CBC-S6-PE,gwastro-mergers-HeeSuk-CompareToPE-Precessing,gwastro-mergers-pe-DoubleSpinBNS-CornishYunes-2014,gwastro-mergers-pe-VitaleALIGO-2014,gwastro-mergers-PNLock-Distinguish-Daniele2015}.
While powerful analytic techniques were developed to estimate the measurement accuracy for nonprecessing binaries
\cite{1992PhRvD..46.5236F,1993PhRvD..47.2198F,1995PhRvD..52..848P,CutlerFlanagan:1994},  
then broadly applied \citeAstroConstraints{}\cite{2005ApJ...629...15H}\citeTideGWBasics{}\cite{2015arXiv150107274B},
for precessing binaries a comparable theoretical tool has remained unavailable.  Instead, the measurement accuracy has been
 evaluated on a case-by-case basis numerically, usually by Bayesian methods that systematically compare the
 data with all possible candidate signals \citeMCMC{}.
The main result of this work is a generalization of the classic analytic approach used to approximate  measurement accuracy
\cite{1995PhRvD..52..848P}  to the case of precessing binaries with a single significant spin undergoing an extended adiabatic,
quasicircular inspiral.  
We restrict to a single significant spin both for convenience -- this limit is well-studied
\cite{ACST,gwastro-SpinTaylorF2-2013,gwastro-nr-imrphenomP} -- and without significant loss of generality -- the smaller
body's spin often has little dynamically significant impact on the angular momentum budget, orbit, or precessional
dynamics ($|\mathbf{S}_2|\le m_2^2 \ll |\mathbf{S}_1,|\mathbf{L}|$), allowing a single-spin model to adequately reproduce the
dynamics and posterior \cite{2015arXiv150606032M}.  
To highlight the broad utility of our approach, we defer  concrete but arbitrary implementation details until we evaluate
numerical results.
Our result is important because it provides the first powerful analytic tool to assess what can be measured using
gravitational waves and why, includes the critical symmetry-breaking effects of spin precession.

\secRO{Inference from GW}
Bayes theorem provides an unambiguous expression for the posterior CBC parameter distribution given instrumental data $\{d\}$; see, e.g, \abbrvILE{} \cite{gwastro-PE-AlternativeArchitectures} for a review.   
The signal and network response to a quasicircular CBC inspiral is characterized by eight intrinsic parameters $\vec{\lambda}$
[$=(m_1,m_2,\mathbf{S}_1,\mathbf{S}_2)$] that uniquely specify the binary's dynamics
 and seven extrinsic parameters $\theta$ [four spacetime coordinates and three
  Euler angles] that specify where, when, and with what orientation the coalescence occurred.   
Each detector responds to an imposed strain as $d_k = n_k + \text{Re}F_k(\unitVector{N})^* h(t+\unitVector{N}\cdot \mathbf{x}_k|\lambda,\theta)$, where $\mathbf{x}_k$ are the
detector positions, $F_k$ are antenna  response functions,  $\unitVector{N}$ is the line of sight to the source,  (a member of
$\vec{\theta}$), $h(t|\lambda,\theta)$ is the strain derived from far-field solution to Einstein's equations for the CBC; and $n_k$ is some random realization of detector noise.
The distribution of stationary gaussian noise $n_k(t)$ in the $k$th detector is completely characterized by its
covariance or power spectrum  $\left<\tilde{n}_k(f)^* \tilde{n}_k(f')\right> = \frac{1}{2} S_k(|f|) \delta(f-f')$.  Let
us define  inner products $\qmstateproduct{\cdot}{\cdot}_k$ on  arbitrary complex functions $a,b$
as $   \qmstateproduct{a}{b}_k \equiv 2  \int_{- \infty}^{\infty} df \frac{a^*(f)b(f)}{S_{k}(|f|)}$.   The log likelihood
ratio  $\Like(\theta,\theta)$ favoring one signal with parameters $\lambda,\theta$ versus no signals is
$2\ln \Like= \sum_k \qmstateproduct{d_k}{d_k}_k -
\qmstateproduct{d_k-h_k(\theta,\lambda)}{d_k-h_k(\theta,\lambda)}$.  If the data is known to contain a signal with
parameters $\Lambda_0\equiv (\lambda_0,\theta_0)$, we will denote the parameters by
$\Lambda=(\vec{\lambda},\vec{\theta})$ and   the
likelihood by $\Like(\Lambda|\Lambda_0, \{n\})$.  The signal amplitude $\rho$ is set by the expected value of the log likelihood ($\rho^2 =2\ln \Like(\Lambda|\Lambda_0,0)$).
Using   15-dimensional posterior distribution  $p_{\rm post}(\lambda,\theta) = \Like p(\theta)p(\lambda)/\int d\lambda
d\theta p(\theta)p(\lambda) \Like(\lambda,\theta)$, the measurement accuracy in some parameter $\lambda_1$ follows from the 90\%
confidence interval derived from  a one-dimensional marginal distribution $p_{\rm post}(\lambda_1)  = \int d\theta
d\lambda_2\ldots d\lambda_8 p_{\rm post}(\lambda,\theta)$.    

While straightforward but expensive numerical techniques exist to estimate the marginal posterior distribution and hence
measurement accuracy for  concrete sources $\Lambda_0$  and noise realizations $n_k$ \citeMCMC{}, an equally
straightforward analytic  approximation to the (average) log likelihood exists at high signal amplitude, the Fisher
matrix
\cite{1992PhRvD..46.5236F,1993PhRvD..47.2198F,1995PhRvD..52..848P,CutlerFlanagan:1994,2007PhRvD..76j4018C,gw-astro-Vallis-Fisher-2007,2008PhRvD..77d2001V}
$\Gamma_{ab}$.  The Fisher information matrix arises in a quadratic-order approximation to the   log-likelihood 
[$
\ln \Like(\Lambda|\Lambda_o) \simeq -\frac{1}{2}\Gamma_{ab}(\Lambda-\Lambda_*)_a (\Lambda-\Lambda_*)_b + \text{const}%
$, with $\Lambda_*$ the location of the noise-realization-dependent maximum]; often depends weakly on the specific noise realization used, particularly at high amplitude; and can be evaluated by
a simple expression involving inner products of derivatives.  For example, for a source directly overhead a network with
equal sensitivity to both polarizations \cite{gwastro-mergers-HeeSuk-FisherMatrixWithAmplitudeCorrections}, the Fisher matrix is
\begin{eqnarray}
\label{eq:Fisher:FiducialDetector}
\Gamma_{ab} = \qmstateproduct{ \frac{dh}{d\lambda_a}} {\frac{dh}{d\lambda_b}}
\end{eqnarray}
The  Fisher matrix can always be evaluated numerically  via direct differentiation \cite{2004PhRvD..70d2001V, 2006PhRvD..74l2001L, 2011PhRvD..84b2002L}, henceforth
denoted Fisher-D.    
For nonprecessing binaries, however, \citet{1995PhRvD..52..848P} %
introduced a powerful analytic
technique, denoted here as Fisher-SPA:  express the signal using a single dominant harmonic,
with a necessarily-simple form; evaluate the fourier transform $\tilde{h}$ via a stationary phase approximation;
thereby evaluate the derivatives $\partial_a h$ \emph{analytically}; and, by reorganizing the necessary integrals
analytically, reduce the evaluation of Eq. (\ref{eq:Fisher:FiducialDetector}) to an analytic expression and a handful of
tabulated integrals.   
Despite its limitations, this method remains the most powerful and widely-used theoretical tool to estimate what can be
measured and why. 
In the remainder of this work, we will review and generalize Fisher-SPA to precessing binaries.

As a matrix in at least 11 dimensions (15 with two precessing spins), the Fisher matrix is both difficult to interpret and highly prone to numerical
instability. 
For nonprecessing binaries,  several studies have demonstrated that the intrinsic and extrinsic posterior distributions largely seperate
\cite{gwastro-mergers-HeeSuk-FisherMatrixWithAmplitudeCorrections,gwastro-mergers-HeeSuk-CompareToPE-Aligned} and that
the intrinsic distribution depends weakly if at all on the specific network geometry.
Hence, to quantitatively assess what can be measured and why for a nonprecessing binary, it suffices to consider a source directly
overhead and optimally aligned with a fiducial detector network
\cite{2013PhRvD..87b4035B,gwastro-mergers-HeeSuk-FisherMatrixWithAmplitudeCorrections,gwastro-mergers-HeeSuk-CompareToPE-Aligned,popsyn-InterpretStarTrack-LIGO-FairhurstOhme-2015}.
In the high-amplitude limit where the likelihood is well-approximated by a gaussian, the event  time $t_c$ and polarization $\psi_c$ can be marginalized out analytically.  
Using this approximation, the log likelihood is approximated using an ``overlap''  $P$
$P(\lambda,\lambda')\equiv
\text{max}_{t\psi}\qmstateproduct{h(\lambda|t_c,\psi_c)}{h(\lambda|0,0)}/||h(\lambda)|| ||h(\lambda')||$ where $||h||\equiv
\sqrt{\qmstateproduct{h}{h}}$, via Eq. (18) of \cite{gwastro-mergers-HeeSuk-FisherMatrixWithAmplitudeCorrections}; the
Fisher matrix arises as a quadratic approximation to the overlap.  
This approach and its relatives, denoted here by Fisher-O and in the literature by overlap, mismatch, or  ambiguity
function methods, has been widely  adopted when analyzing numerical simulations
\cite{2008PhRvD..78l4020L,2009JPhCS.189a2024M,2009PhRvD..79h4025H,2010PhRvD..82l4052H,2011PhRvD..84j4017P}  and circumvents the numerical challenges that plague brute-force
11-dimensional calculations. 

Even for precessing binaries, several studies have suggested that the four spacetime coordinates decouple from intrinsic
parameters and the binary's three Euler angles
\cite{gwastro-mergers-HeeSuk-CompareToPE-Precessing,gw-astro-Tides-MontanaWithPrecessionViaTheirModel-2015}.  To
simplify subsequent analytic calculations, following prior work \cite{gwastro-mergers-HeeSuk-FisherMatrixWithAmplitudeCorrections,gwastro-mergers-HeeSuk-CompareToPE-Aligned} we will therefore adopt the ansatz that the source can be assumed directly
overhead a network with equal sensitivity to both polarizations.

\secRO{Stationary-phase approximation} The outgoing GW signal $h(t,\hat{n},\lambda)$ is modeled using a stationary phase approximation of the
leading-order (corotating) quadrupole emission, assuming a single significant spin.  Specifically, adopting the
conventions of \cite{gwastro-SpinTaylorF2-2013} and \abbrvILE, we express the strain for  a source with intrinsic parameters $\lambda$ as a spin-weighted
spherical harmonic expansion $h(t|\lambda,\theta)=h_+-i h_\times = e^{-2i\psi_J}(M/d_L) \sum_{lm} h_{lm}(t-t_c|\lambda)\Y{-2}(\unitVec{n})$, relative to a cartesian frame $\unitVec{z}' = \unitVec{J},\unitVec{x}',\unitVec{y}'$ defined by the total
angular momentum $\mathbf{J}$, where $d_L$ is the luminosity distance to the source, $M$ is the binary mass,
$\unitVec{n}$ is the propagation direction away from the source, $t_c$ is the coalescence time, and
$\psi_J$ is the angle of  $\mathbf{J}$ on the plane of the sky.   
Within the post-Newtonian approximation, both the amplitude and phase of these
functions $h_{lm}(t)$ and  their (stationary-phase) fourier transforms are slowy-varying and analytically-tractable
functions \cite{gw-astro-PN-Comparison-AlessandraSathya2009}.
For a nonprecessing binary, the sum is dominated by a single pair of complex-conjugate terms
$h_{22}=h_{2,-2}^*=|h_{22}|e^{-2i\Phi}$, enabling efficient calculation of Eq. (\ref{eq:Fisher:FiducialDetector})
\cite{1995PhRvD..52..848P} for an optimally-aligned source directly overhead a single detctor (Fisher-O).

A generic quasicircular binary will orbit, precess, and inspiral on three well-seperated timescales $1/f_{\rm orb}$,
$t_{prec}$, and $t_{\rm insp}$.  For this reason, to a good approximation, the gravitational radiation from a
precessing, inspiralling binary \cite{WillWiseman:1996,lrr-Blanchet-PN}  can be approximated as if from an instantaneously
nonprecessing binary: $h_{lm}(t|\lambda)=  \sum_{\bar{m}} h_{l\bar{m}}^{(\rm C)}(t,\lambda) D^l_{m\bar{m}}(R(t))$, where
$R(t$) is  a minimal rotation transforming $\unitVector{z}$ into $\unitVector{L}$  \cite{gwastro-mergers-nr-Alignment-BoyleHarald-2011,gwastro-SpinTaylorF2-2013,gwastro-nr-imrphenomP}, where
$h_{lm}^{(C)}$ are available in the literature
\cite{WillWiseman:1996,gw-astro-mergers-approximations-SpinningPNHigherHarmonics,2014arXiv1409.4431B} in terms of the
spins, the orbital phase $\Phi_{orb}$, and a post-Newtonian expansion parameter $v=(Md\Phi_{orb}/dt)^{1/3}$.   
The quantities appearing in these expressions ($\mathbf{L},\mathbf{S}_i,\Phi_{orb orb},v$) are determined by
 post-Newtonian approximations that  prescribe the evolution of
both spins  [$\partial_t \mathbf{S}_i = \mathbf{\Omega}_i\times \mathbf{S}_i$\ and the orbital phase 
[$\frac{dv}{dt}= - \frac{{\cal F} + \dot{M}}{E'(v)}$],
\cite{LIGO-nr-format-2007,gw-astro-PN-Comparison-AlessandraSathya2009,gwastro-AndySpinModelsDiffer-2013}, 
where $\dot{M}$ is the rate at which the black holes' mass changes  \cite{2001PhRvD..64j4020A, 2013PhRvD..87d4022C},
henceforth neglected;
${\cal F}(v)$ is the rate at which energy is radiated to infinity; and $E(v)$ is the energy of an instantaneously
quasicircular orbit; all of which are provided in the literature.  
At leading amplitude order, corotating-frame strain satisfies $h_{lm}^{(C)}(t|\lambda)= |h_{lm}(t)| \exp (- i
\Phi_{orb})$, where $|h_{lm}|$ is a slowly-varying function of $v$; substituting this form into the general expression implies
\begin{align}
\label{eq:hlm:SimpleExpansion}
h_{lm}(t|\lambda)= \sum_{\bar{m}}  e^{-i \bar{m}(\Phi_{\rm orb}+\gamma)} e^{-im\alpha}  d^l_{m\bar{m}}(\beta) |h_{lm}(v)|
\end{align}
In this expression, we have expanded the rotation $R(t)$ using Euler angles, set by the orbital angular momentum direction  $\hat{\mathbf{L}}$  expanded relative to the (assumed fixed) total angular
momentum direction $\hat{\mathbf{J}}$ as $\hat{\mathbf{L}}=  \sin \beta_{JL} \cos \alpha_{JL} \hat{x}' + \sin \beta_{JL} \sin \alpha_{JL} \hat{y}' + \cos
\beta_{JL} \hat{\mathbf{J}} $;  the remaining Euler angle $\gamma =  - \int  \cos \beta_{JL} d\alpha_{JL}$.  
In this work, we restrict to the leading-order gravitational-wave quadrupole $h_{2,\pm 2}^{(C)} =-8\sqrt{\pi/5} \eta
v^2 \exp( \mp i \Phi_{\rm orb})$, so this sum has only two terms.

For a binary with a single dynamically significant spin, the spin-precession equations imply that  the orbital angular
momentum precesses simply around the total angular momentum: $\beta$ changes slowly, on the inspiral timescale, while
$\alpha$ and $\gamma$ evolve on the precessional timescale \cite{ACST,gwastro-SpinTaylorF2-2013}.   

Because of seperation of timescales, because of the simple form of Eq. (\ref{eq:hlm:SimpleExpansion}), and particularly
because the phase terms $\bar{m}(\Phi+\gamma)+m\alpha$ are monotonic and well-behaved, the
stationary-phase approximation to the fourier transform $\tilde{h}_{lm}(\omega) = \int dt h_{lm} \exp{i \omega t}$ can
be carried out term by term  \cite{gwastro-SpinTaylorF2-2013}.  For each term, the stationary-phase condition defines an
$m,\bar{m}$-dependent time-frequency trajectory $\tau_{m\bar{m}}(\omega)$ set by solving
\begin{eqnarray}
\omega \equiv \bar{m} (\dot\Phi_{\rm orb}-\dot{\alpha}\cos \beta_{JL}) + m \dot \alpha
\end{eqnarray}
or, equivalently,  $v =( M \Phi'_{\rm orb})^{1/3} = 
  \left[
  M \frac{\omega - \bar{m}\dot{\gamma}(\tau_{m\bar{m}}) - m \dot{\alpha}(\tau_{m\bar{m}})}{\bar{m}}
  \right]^{1/3}$.
Using this time-frequency relationship to evaluate $\Psi_{m,\bar{m}}\equiv \omega t-\bar{m}(\Phi_{\rm orb}+
\zeta)-m\alpha $ 
and the slowly-varing coefficients in each term in Eq. (\ref{eq:hlm:SimpleExpansion}), the stationary-phase approximation is
\begin{widetext}\begin{align}
\tilde{h}_{lm}(\omega)
& \equiv  \sum_{\bar{m}}
\begin{cases}
\frac{
  d^l_{m\bar{m}}(\beta(\tau_{m\bar{m}}(\omega))) |h_{l\bar{m}}^{\rm ROT}(\tau_{m\bar{m}}(\omega))|e^{i\Psi_{m\bar{m}}(\omega)}
}{
\sqrt{i (\bar{m}(\Phi''_{\rm orb}+ \zeta'')+m \alpha'')/2\pi}} 
  & \bar{m}\omega>0 \\
0 & \bar{m}\omega <0
\end{cases}
\label{eq:def:SPA}
\end{align}
\end{widetext}
This expression for the SPA had been previously derived in the restricted PN approximation by precisely
this method  \cite{gwastro-SpinTaylorF2-2013}, for simplicity neglecting the distinct time-frequency trajectories
implied by $\tau_{m,\bar{m}}$.  

\secRO{Inner products via a time-frequency ansatz}
Most of the terms in Eq. (\ref{eq:hlm:SimpleExpansion}) are mutually orthogonal.  For example, the modes with
$\bar{m}>0$  and $\bar{m}<0$ have different helicity and have almost no overlap, indepdendent of the precession state
\citeComplexOverlap{}.  Additionally, for binaries with more than a  few precession
cycles in band  (see, e.g., \cite{gw-astro-SpinAlignedLundgren-FragmentA-Theory} for suitable conditions on
$|\mathbf{S}_1|, m_1,m_2$),
each term in Eq. (\ref{eq:hlm:SimpleExpansion}) is associated with a unique time-frequency trajectory and hence is
orthogonal to all others.  
Using this ansatz, the inner product $\qmstateproduct{h(\Lambda)}{h(\Lambda')}$ for $\Lambda\simeq \Lambda'$ can be
approximated using a sum over 10 terms: 5 for each of the $l=2$ modes $m=-2,-1,\ldots 2$ and, for each mode, one term
for each helicity.

By contrast, to evaluate the overlap of terms that are \emph{not} orthogonal, the specific time-frequency trajectory has
relatively little impact.  We therefore approximate  $\tau_{m\bar{m}}\simeq  \tau_{0\bar{m}}$ henceforth.

\secRO{Fisher matrix} We now use the time-frequency ansatz and the restricted PN approximation to evaluate a Fisher matrix for a source directly overhead an
idealized network of two interferometers oriented to have equal sensitivity to both polarizations
[Eq. (\ref{eq:Fisher:FiducialDetector})].  Because of the time-frequency ansatz, the overall Fisher matrix is a weighted
sum of
10 individual Fisher matricies, associated with each harmonic:
\begin{widetext}\begin{align}
\Gamma_{ab} &= \sum_{m=2}^{2} \sum_{s=\pm 1} \rho_{2ms}^2 \hat{\Gamma}_{ab}^{ms} \\
\rho^2_{2ms}& \equiv  |\Y{-2}_{2m}(\theta_{JN}) d^2_{m,2s}(\beta)|^2  \int_0^{\infty} \frac{df}{S_h(f)}\frac{4(\pi
  \mc^2)^2}{3 d_L^2} (\pi \mc f)^{-7/3} \\
\label{eq:HatGamma}
\hat{\Gamma}_{ab}^{(ms)}&=
\frac{\int_0^{\infty} \frac{df}{S_h(f)} (\pi \mc f)^{-7/3} 
\partial_a(\Psi_2 - 2\zeta -m s \alpha)\partial_b(\Psi_2 -2\zeta-ms \alpha)
}{
\int_0^{\infty} \frac{df}{S_h(f)} (\pi \mc f)^{-7/3}}
\end{align}
\end{widetext}
In this expression,  $\Psi_2 = \omega t - 2 \Phi_{orb}$ is the stationary-phase approximation derived via 
$\omega =2 \dot{\Phi_{orb}}$.     The weights $\rho_{ms}$ are associated with the relative contribution of
each model $m$ and sign $s$ to the detected amplitude, along this line of sight.  The 10 individual Fisher matrices $\hat{\Gamma}_{ab}$
reflect the Fisher matrix implied by a \emph{single harmonic}, with a modified phase versus time to reflect that
harmonic's precession-induced \emph{secular} phase change; each one reduces trivially to the well-known nonprecessing
Fisher matrix in the absence of precession.

Because measurements often cannot tightly constrain all parameters, whether computed directly (Fisher-O) or via our
approximation, the Fisher matrix is often degenerate, particularly in phase angles.  Following prior work, when evaluating the Fisher matrix to
produce our final numerical results, we adopt a very weak (regularizing) prior to break degeneracy: $\Gamma_{final} = \Gamma+K$ where $K$
reflects a multivariate gaussian distribution with standard deviation $2\pi$  in phase angles, $1 M_\odot$ in total
mass, 1 second in time, and $1/4$ in mass ratio. 
\begin{figure}
\includegraphics{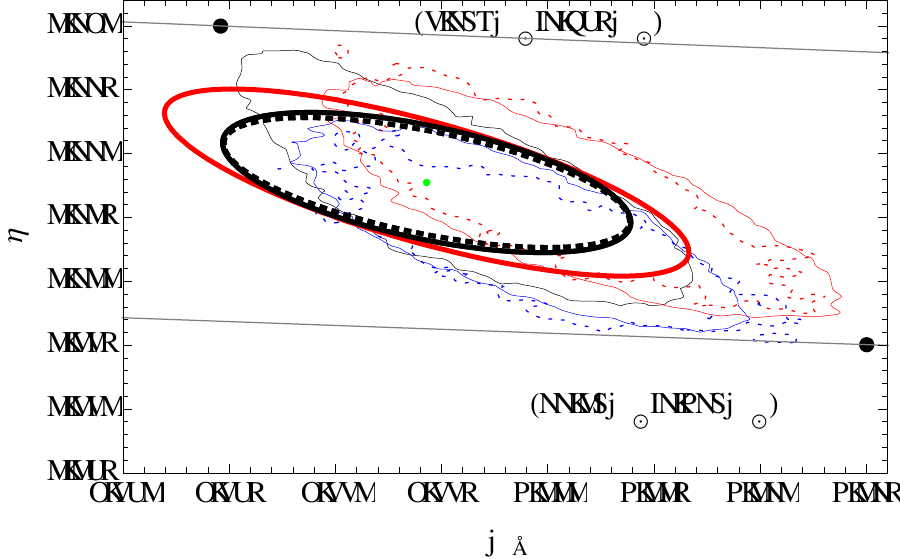}
\caption{\label{fig:ComparisonPrecessing}\textbf{Comparison}: Comparison between marginalized Fisher matrix in
  $\mc,\eta$ for a precessing binary evaluated using the ansatz in this paper (red solid curve: Fisher-SPA); via full
 15-dimensional  MCMC (thin  curves); and via a quadratic approximation to the 7-dimensional overlap (Fisher-O).
 The latter two results, previously presented \citet{gwastro-mergers-HeeSuk-CompareToPE-Precessing}, used with permission.
}
\end{figure}

\secRO{Implementation and results}
Figure \ref{fig:ComparisonPrecessing} shows a comparison beteween our approximation; a Fisher-O approximation; and a
full  MCMC posterior distribution.  
In our calculation, to minimize superfluous differences associated with uncontrolled post-Newtonian remainders, we
evaluated the phase functions $\Psi_2(v),\gamma(v),\alpha(v)$ needed for our Fisher-SPA approximation
\emph{numerically},  using precisely the same post-Newtonian evolution model adopted in the other
calculations shown \cite{gwastro-mergers-HeeSuk-CompareToPE-Precessing} to evaluate  $t(v)$ and $\Phi_2(t),\alpha(t),\gamma(t)$; see that work for the specific post-Newtonian
approximation used.  
Despite not including the prior $p(\lambda)$ and despite adopting highly simplifying assumptions, our expression shows
good agreement with the multidimensional posterior and previous numerical estimates of the marginalized Fisher-O matrix.
Though not shown here, similar results follow by using the explicit expressions for $\Psi_2(v),\gamma(v),\alpha(v)$ available
the literature \cite{gwastro-SpinTaylorF2-2013}.

\secRO{Future directions}
While only approximating the results of detailed Bayesian parameter estimation
\cite{2014PhRvD..89d2004G,gwastro-skyloc-Sidery2013,gwastro-em-First2Years-2014}, Fisher-matrix calculations provide a
powerful and analytically-tractable tool to assess what can be measured and why.  Extending the Fisher-SPA method to
include a single precessing spin will help rapidly interpret of real gravitational wave data, via improved
 methods to explore the model space and interpret the posterior; 
assess the impact of  systematic errors from the waveform model; 
quantify the accuracy to which tidal effects and   modifications of general relativity can be detected; and otherwise
understand what can be measured and why.

Further investigation is needed to generalize our approach to account for a second significant spin, using
recently-developed analytic solutions
\cite{gwastro-mergers-PNLock-Morphology-Kesden2014,gwastro-mergers-PNLock-PRLFollowup,gwastro-mergers-PNLock-Distinguish-Daniele2015};
and to perform a large-scale comparison between our calculations and detailed Bayesian parameter estimates.
Finally, to facilitate the immediate use of our approach and enhance its similarity to prior work, we have adopted
extremely simple assumptions (e.g., the neglect of $\tau_{m,\bar{m}}$; the neglect of all but 10 overlaps; and the
restricted PN expansion).  These
assumptions can easily be relaxed if more detailed calculations are required,  since the simple form of
Eq. (\ref{eq:def:SPA}) insures a theoretically-tractable analysis.

\secRO{Acknowledgements}
ROS acknowledges support from   NSF PHY-1505629 and PHY-0970074.

\bibliography{paperexport}

\end{document}